\documentclass[useAMS,usenatbib]{mn2e}
\usepackage{natbib}
\usepackage{epsfig}
\usepackage{amsbsy}
\usepackage{hyperref}
\providecommand{\adsurl}[1]{\href{#1}{ADS}}

\newcommand{\lya}{Lyman-$\alpha$~}

\newcommand{\be}{\begin{equation}}
\newcommand{\ee}{\end{equation}}
\newcommand{\ba}{\begin{eqnarray}}
\newcommand{\ea}{\end{eqnarray}}
\newcommand{\brr}{\begin{array}}

\newcommand{\err}{\end{array}}
\newcommand{\bc}{\begin{center}}
\newcommand{\ec}{\end{center}}

\newcommand{\hm}{\,h^{-1}{\rm Mpc}}

\newcommand{\mincir}{\raise
  -2.truept\hbox{\rlap{\hbox{$\sim$}}\raise5.truept \hbox{$<$}\ }}
\newcommand{\magcir}{\raise
  -2.truept\hbox{\rlap{\hbox{$\sim$}}\raise5.truept \hbox{$>$}\ }}
\newcommand{\siml}{\raise
  -2.truept\hbox{\rlap{\hbox{$\sim$}}\raise5.truept \hbox{$<$}\ }}
\newcommand{\simg}{\raise
  -2.truept\hbox{\rlap{\hbox{$\sim$}}\raise5.truept \hbox{$>$}\ }}

\DeclareMathAlphabet{\mathsc}{OT1}{cmr}{m}{sc}
\def\testbx{bx}%
\DeclareRobustCommand{\ion}[2]{%
\relax\ifmmode
\ifx\testbx\f@series
{\mathbf{#1\,\mathsc{#2}}}\else
{\mathrm{#1\,\mathsc{#2}}}\fi
\else\textup{#1\,{\mdseries\textsc{#2}}}%
\fi}
\title[The evolution of a pre-heated Intergalactic Medium]
{The evolution of a pre-heated Intergalactic Medium}

\author[Borgani \& Viel] 
{S.  Borgani$^{1,2,3}$ \&
M. Viel$^{2,3}$\\
$^1$ Dipartimento di Astronomia dell'Universit\`a di Trieste,  Via
G.B. Tiepolo 11, I-34131 Trieste, Italy\\
$^2$ INAF - Osservatorio Astronomico di Trieste, Via G.B. Tiepolo 11,
I-34131 Trieste, Italy (borgani,viel@oats.inaf.it)\\
$^3$ INFN - National Institute for Nuclear Physics, Via Valerio 2,
I-34127 Trieste, Italy\\
\\}

\begin{document}
\maketitle
\begin{abstract}
  We analyse the evolution of the Intergalactic Medium (IGM) by means
  of an extended set of large box size hydrodynamical simulations
  which include pre--heating. We focus on the properties of the $z\sim
  2$ \lya forest and on the population of clusters and groups of
  galaxies at $z=0$. We investigate the distribution of voids in the
  \lya flux  and the entropy--temperature relation of
  galaxy groups, comparing the simulation results to recent data from
  high-resolution quasar spectra and from X--ray
  observations. Pre--heating is included through a simple
  phenomenological prescription, in which at $z=4$ the entropy of all
  gas particles, whose overdensity exceeds a threshold value
  $\delta_{\rm h}$ is increased to a minimum value $K_{\rm fl}$. While
  the entropy level observed in the central regions of galaxy groups
  requires a fairly strong pre--heating, with $K_{\rm fl}>100$ keV
  cm$^2$, the void statistics of the \lya forest impose that this
  pre--heating should take place only in relatively high--density
  regions, $\delta_{\rm h}\magcir 30$, in order not to destroy the
  cold filaments that give rise to the forest. We conclude that any
  injection of non--gravitational energy in the diffuse baryons should
  avoid low--density regions at high redshift and/or take place at
  relatively low redshift $z\mincir 1$.
\end{abstract}

\begin{keywords}
cosmology: observations -- methods: numerical -- intergalactic
medium
\end{keywords}

\section{Introduction}
Observations in the X--ray band of the hot intra--cluster and
intra--group medium, along with observations of the absorption features
in the spectra of distant quasars (QSOs) from the intervening intergalactic
medium (IGM) offer powerful means of tracing the evolution of diffuse
cosmic baryons in different regimes. While X--ray observations of
galaxy systems trace the high--density baryons in the low--redshift
$z\mincir 1$ Universe \citep[e.g.][]{voit05}, data on the \lya forest
convey information on baryons around the mean cosmic density at
$z\magcir 2$ \citep[e.g.][]{meiksin07}. As such, these different
observational techniques complement each other in the reconstruction
of the cosmic cycle of baryons.

One of the standard results from the X--ray observations of groups and
clusters of galaxies is that the hot gas in the central regions of
low--temperature systems has an entropy level higher than predicted by
the gravitational process of accretion shocks
\citep[e.g.][]{TN01}. The commonly accepted explanation for this is
that radiative cooling and heating from some feedback energy
source should be the main mechanisms responsible for setting the hot
intra--cluster medium on a relatively high adiabat, while preventing
overcooling from converting into stars a unrealistically large
fraction of baryons \citep[][]{voit05}. In order to test these
predictions, hydrodynamical simulations including some form of extra
gas heating have been analysed by a number of authors \citep[][
for a recent review]{borgani08}. The general result from these
simulations is that injecting $\sim 1$ keV per gas
particle is effective in increasing the central gas entropy to a level
consistent with the observed one, while the question remains open as
to what astrophysical mechanism should be responsible for this
high--redshift heating. Clearly, any diffuse IGM heating at high
redshift is expected to leave an imprint on its observational
properties. For instance, increasing the gas temperature inside the
filaments permeated by the IGM should significantly alter the
statistical properties of the \lya forest. Indeed, \cite{shang}
recently suggested that the pre--heating required to reproduce 
X--ray cluster observations should leave the
imprint of bubbles of ionised gas at high redshift. These authors
analysed a sample of SDSS QSO spectra searching for large voids in the
\lya forest associated to these bubbles. From the lack of detection of
these voids, they concluded that any pre--heating should only involve
IGM at a density higher than the mean one as traced by the \lya
forest.

In this Letter we present a combined analysis of the void statistics
in the \lya forest at $z=2.2$ and of the entropy level of the hot
gas within galaxy groups at $z=0$, from a set of pre--heated
cosmological hydrodynamical simulations within a fairly large
cosmological box. The analysis of the void statistics in the \lya
forest will be carried out by using the same procedure presented by
\cite{viel08voids}, where a description of the reference observational
data set is also provided. As for the entropy of the intra-group gas,
we will compare simulation predictions to the recently published
results by \cite{sun08}, who analysed {\small{CHANDRA}} data for an
extended set of nearby groups. By changing both the strength of the
pre--heating and the minimum overdensity at which such a pre--heating
takes place, we aim at better exploiting the complementary
information placed by high-density hot gas in nearby groups/clusters
and by the lower--density IGM associated to the high--redshift \lya
forest.

\section{Hydrodynamical simulations}
\begin{figure}
\begin{center}
\includegraphics[width=9cm, height=9cm]{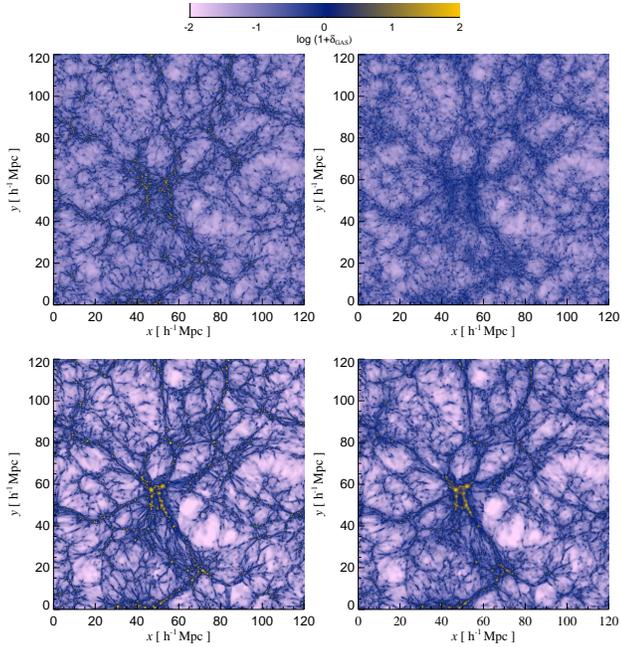}
\end{center}
\vspace{-0.5truecm}
\caption{The different panels show the gas overdensity in a slice of 8
  comoving $\hm$ for the reference simulation, with no pre--heating
  (left panels) and for the (300,10) simulation (entropy floor of 300
  keV\,cm$^2$ for $\delta_{\rm GAS}>10$, right panels). Top and bottom
  panels refer to the snapshots at $z=2.2$ and at $z=0$, respectively.}
\label{fig1}
\end{figure}

We use simulations carried out with the Tree-SPH {\small {GADGET-2}}
code \citep{springel}. These simulations include radiative cooling and
heating from a uniform evolving UV background (see
  \cite{bolton05} for details), for a primordial mix of hydrogen and
helium. Since no self--consistent model for chemical enrichment
  is treated in these simulations, we decided not to include the
  effect of metals in the computation of the cooling function. The
star formation criterion simply converts in collisionless stars all
the gas particles whose temperature falls below $10^5$ K and whose
density contrast is larger than 1000.  More details can be found in
\cite{viel04}. The cosmological model assumed for our simulations
corresponds to a $\Lambda$CDM Universe with $\Omega_{\rm m }=0.26,\
\Omega_{\Lambda}=0.74,\ \Omega_{\rm b }=0.0463$, $n_{\rm s}=0.95$, and
$H_0 = 72$ km s$^{-1}$ Mpc$^{-1}$ and $\sigma_8=0.85$, in agreement
with the latest results of cosmological parameters derived from CMB
and large--scale structure observables \citep{lesg,komatsu08}. We have
used $2\times 400^3$ dark matter and gas particles in a $120\ h^{-1}$
comoving Mpc box for our purposes. The gravitational softening was set
to 15 $h^{-1}$ kpc in comoving units for all the particles.  Even if
these simulations barely resolve the Jeans length, \cite{viel08voids}
showed that the statistics of voids have numerically converged once
the simulated spectra have been smoothed over a scale of 1$\hm$
comoving. With this mass resolution, the smallest groups in the
observational data set by \cite{sun08} which have $M_{500}\simeq
10^{13}\hm$, are resolved with more than 5000 particles.

Non--gravitational heating is included following a standard purely
phenomenological model of pre--heating, based on imposing a minimum
entropy floor, $K_{\rm fl}$ to all the gas particles that, at a given
heating redshift $z_{\rm h}$, have density contrast above a given
threshold $\delta_{\rm h}$. We use the definition of
entropy which is standard in the study of the intra--cluster medium,
namely, $K=n_e/T$, where $n_e$ is the electron number density and $T$
the gas temperature. Similar schemes to study the
effect of pre--heating to the X--ray properties of galaxy clusters
have been used by several authors (e.g.
\citealt{bialek01,muanwong02,tornatore2003,borgani05,younger07}). The
procedure adopted to pre--heat the gas is the following: at $z_{\rm
  h}=4$ we select from the reference run, which does not include any
pre--heating, all the gas particles having $\delta>\delta_{\rm h}$ and
entropy $K<K_{\rm fl}$. The internal energy of these particles is then
increased so that their entropy match the floor value. In this
selection, we always exclude cold and dense star particles, with
$\delta>500$ and $T<3\times 10^4$ K, whose low initial entropy would
require an extremely strong and ad--hoc heating. Such cold and
  dense gas is contained within galaxy--sized halos, where star
  formation takes place \citep[e.g.,][]{kay00}. Therefore, imposing
  the above limits on density and temperature of the gas to be
  pre-heated amounts to protect the cold gas content of galaxies, which
  should be rather affected by some local feedback mechanims (e.g.,
  associated to SN explosions). 
We refer to \cite{voit2005} for an overview of the role
  of pre-heating in modelling the thermodynamics of the intra-group
  medium.

We choose four different values of the entropy floors: $K_{\rm
  fl}=10,50,100,300\,{\rm keV\,cm^2}$; and 3 overdensity thresholds:
$\delta_{\rm h}=-1,3,10$. Only for the case with $K_{\rm fl}=300\,{\rm
  keV\,cm^2}$ we also consider the case with a higher density
threshold for heating, $\delta_{\rm h}=30$. While the lowest
$\delta_{\rm h}$ corresponds to heating all the gas, the highest value
is comparable to that expected at the boundary of a virialized
structure. In this way, we run a total of 14 simulations (13 different
pre--heated runs and the default run with no pre--heating). In the
following, we will show results only for a subset of the whole
simulation set. The different simulations will be labelled by the
tuple of values of entropy floor (in keV\,cm$^2$) and of overdensity
threshold for heating. The reference run will be indicated with the
label {\em REF}.  All the simulations are analysed at $z=2.2$ to
compute the \lya forest statistics and at $z=0$ to compute the
entropy--temperature relation of the galaxy groups.

The sample of high-resolution, high signal-to-noise QSO spectra is
described in \cite{tkim} and has been already used to compute several
flux statistics \citep{bolton08,viel04,viel08voids}. Many systematic
effects have been carefully addressed including metal lines
contamination and continuum fitting errors. The median redshift of the
sample is at $z=2.2$.  From this data set we extracted the cumulative
distribution of voids in the transmitted flux, i.e. connected regions
with flux above the mean level, which at $z\sim 2$ appear to trace
reasonably well underdense regions.
\begin{figure*}
\begin{center}
\includegraphics[width=14cm,height=6.cm]{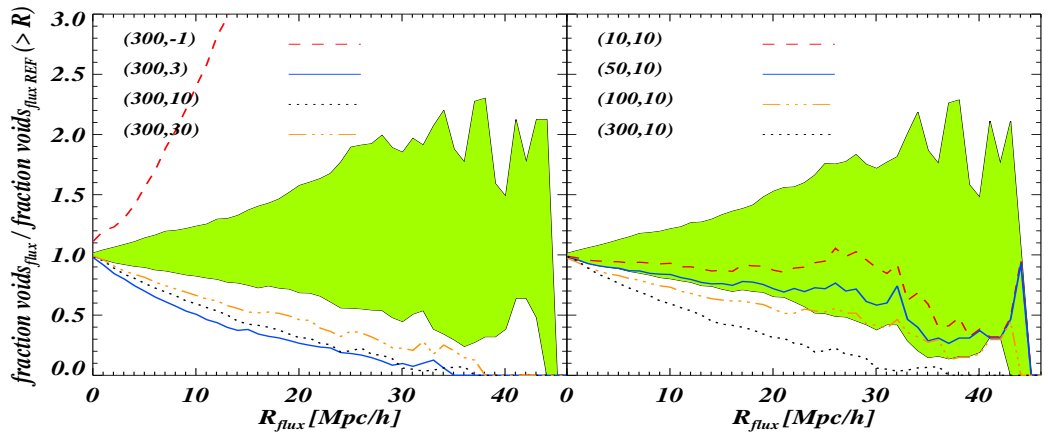}
\end{center}
\caption{The ratio between \lya flux voids larger than $R_{flux}$ (in
  comoving $\hm$) of pre--heated and the no--heating (REF) case. In
  the left panel we show the effect of a different gas density
  threshold having fixed the entropy floor for the models (300,-1),
  (300,3), (300,10) and (300,30). In the right panel we show the
  effect of a different value for the entropy floor having fixed the
  density threshold for the models (10,10), (50,10), (100,10),
  (300,10). The shaded area in both panels could be considered as the
  $1-\sigma$ estimate of the observational statistical error (see
  text).}
\label{fig2}
\end{figure*}

As for the {\small{CHANDRA}} sample of galaxy groups, it contains 40
objects selected by \cite{sun08} so that gas properties can be derived
out to $r_{2500}$ for all of them ($r_\Delta$ is the radius
encompassing an average density of $\Delta \times\rho_{\rm c}$, being
$\rho_{\rm c}$ the cosmic critical density). \cite{sun08} analysed this
sample to derive the scaling relation between entropy and temperature
at different overdensities and found an excess of entropy at
$r_{500}$, with respect to the baseline value calibrated by
\cite{voit05} from non--radiative hydrodynamical simulations of
clusters, although this excess is smaller than suggested by previous
analyses \citep[e.g.][]{ponman03,pratt05,piffaretti05}. Moreover,
they also found that the excess entropy is larger at $r_{2500}$, thus
confirming that any non--gravitational process has a larger effect in
the central regions of groups.

\section{Results}
In Figure \ref{fig1} we show a projected slice whose thickness is 8
comoving $\hm$ of the gas distribution at $z=2.2$ and $z=0$, for the
default run and for the (300,10) run. Even by adding this strong
pre--heating, at $z=2.2$ the skeleton of the cosmic web at densities
around the mean is still preserved.  The main differences arise in
dense structures, whose filling factor is small. At $z=0$ the effect
of pre--heating is quite visible as well: the mildly non-linear cosmic
web evolves and gives rise to clusters of galaxies and galaxy
groups. These very non-linear structures tend to be puffier and
smoother compared to the reference case, with a suppression of the
number of small halos that trace the filaments.

\subsection{Void statistics in the \lya forest}
From the snapshots at $z=2.2$ we extract a mock set of 1000 \lya QSO
spectra in random directions. Then, we smooth the spectra over a scale
of 1 comoving $\hm$, to be less sensitive to small structures at and
around the Jeans length that might not be properly resolved by our
simulations. We then compute the number of voids in the flux
distribution, having size larger than a given value and compare the
pre--heated runs with the default one. As shown by \cite{viel08voids},
the reference run provides results in excellent agreement with the
data. Results are presented in Figure \ref{fig2}, where the shaded
area indicates the uncertainty in the observed mean flux level, which
sets the criterion for the selection of voids. Among all the possible
uncertainties in the astrophysical and cosmological parameters, this
error has the largest effect on the void statistics (see
\citealt{viel08voids} for more details). Therefore, we take it as a
rough estimate of the error. We remind that the total error budget
must also take into account the contributions from all the other
parameters and it is larger than the one shown here. For example, a
$\sim 3$ times higher temperature of the IGM at the mean density would
boost the number of 30 comoving $\hm$ voids up to a factor 1.5,
bringing some of the models in better agreement with the
observations. Basically, the error bars represented by the shaded area
do not take into account different thermal histories for the low
density IGM and/or different cosmological scenarios (warm dark matter
or extra power at intermediate scales) that are more extensively
discussed in \cite{viel08voids}.

In the left panel, we address the role of a different overdensity
threshold for heating, keeping fixed the entropy floor at 300 keV
cm$^2$. Clearly, heating up the whole IGM (i.e. $\delta_{\rm h}=-1$)
substantially increases the number of voids since the neutral fraction
will be reduced by the very large amount of heating. If we increase
the overdensity threshold to $\delta_h=3$ and 10, then the void fraction is
in better agreement with the default case, as expected, but there is a
tendency to under-predict the number of large void regions. This
result might seem counter--intuitive, since a heating of the IGM
should reduce the hydrogen neutral fraction. However, we found that
the gas in the relatively low density environments of the pre--heated
runs is denser compared to the default run and thereby carries with it
a larger neutral hydrogen fraction. The opposite trend takes place in
overdense regions where the default simulation is denser than the
corresponding pre--heated ones. At low redshifts the volume
  filling factor of underdense (overdense) regions gets larger
  (smaller) in all the models but the overall effect is a reduction of
  the size of large voids compared to the REF case since in the
  pre--heated runs the voids are less empty.  The (300,30) is closer
to the default case than the (300,10) owing to the decrease of the
volume filling factor of the heated regions with increasing
$\delta_{\rm c}$.

In the right panel of Figure \ref{fig2} we show the effect of changing
the entropy floor, while keeping the overdensity threshold for heating
fixed at $\delta_{\rm h}=10$. Increasing the entropy floor has the effect of
lowering the fraction of large voids: at $z=4$ the heating of dense
gas particles at $\delta > 10$ makes them leave quickly their halos
and reach the low density IGM: these particles have usually a larger
fraction of neutral hydrogen that can cause absorptions in the mock
spectra that are extracted at $z=2.2$.

\begin{figure*}
\begin{center}
\includegraphics[width=14.5cm, height=12.5cm]{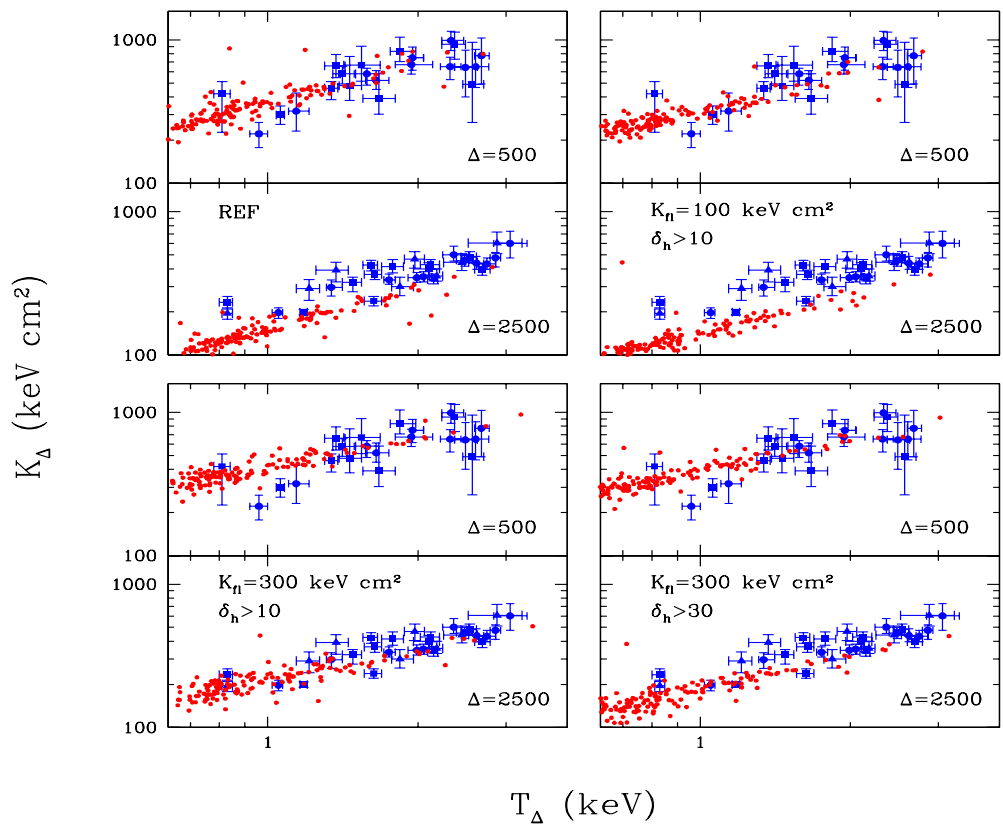}
\end{center}
\vspace{-1.truecm}
\caption{The relation between entropy and temperature for simulated
  groups at $z=0$ (small circles) and for the {\small{CHANDRA}}
  observational of nearby groups analysed by \protect\citep{sun08}
  (points with error bars). The four panels correspond to the REF run
  (with no pre--heating; top left), to the (100,10) run (top right), to
  the (300,10) run (bottom left) and to the (300,30) run (bottom
  right). The top and the bottom parts of each panel refer to the
  $K_\Delta$--$T_\Delta$ relation for $\Delta=500$ and $\Delta=2500$,
  respectively.}
\label{fig3}
\end{figure*}

\subsection{The entropy--temperature relation of groups}
At $z=0$ we identify galaxy groups in our simulations by running a
friends--of--friends algorithm with $b=0.15$ for the linking length in
units of the mean dark matter (DM) interparticle separation. The
centre of each group is then identified with the position of the DM
particle having the minimum value of the gravitational potential.
Following \cite{sun08}, we compute for each group the temperature
$T_\Delta$ within $r_{\Delta}$ (for $\Delta=500$ and 2500) by
excluding the core region within $0.15\,r_{500}$. Then we measure the
profiles of the electron number density, $n_{\rm e}(r)$, and of the
gas temperature, $T(r)$. The values of entropy $K_\Delta$ at
$r_\Delta$ are then computed as $K_{\Delta}=T_{\Delta}/n_{\rm
  e,\Delta}^{2/3}$. $X$--ray temperatures are computed following
the prescription by \cite{vikhlinin06}, which represents an extension
to low--temperature systems ($T_X< 3$ keV, relevant for our analysis)
of the spectroscopic--like temperature, originally
introduced by \cite{mazzotta2004}.

In Figure \ref{fig3} we compare the relation between entropy and
temperature for our simulated groups to the observational data by
\cite{sun08}, for four of our simulated boxes. As for the case with no
pre--heating (upper left panel), a satisfactory agreement with data is
obtained at $r_{500}$, while simulated groups have a too low level of
entropy at $r_{2500}$. This result is in line with the comparison
performed by \cite{sun08} between their observational results and the
best--fit relation from the hydrodynamical simulations by
\cite{nagai07}. These simulations include radiative cooling, star
formation and a rather inefficient form of feedback, and therefore are
expected to provide similar results to the reference (non
pre--heated) run. Clearly, the enntropy level at $r_{2500}$ is
not the only problem suffered by a radiative run without extra
heating. Indeed, in this simulation overcooling causes about 40--50
per cent of the baryons within $r_{500}$ to be converted in stars,
with a decreasing trend with the system temperature. This fraction,
which is a lower limit owing to the resolution dependence of the
cooling efficiency \citep[e.g.][]{balogh01,borgani06}, is in excess
with respect to observational estimates
\citep[e.g.][]{gonzalez07}. Therefore, while cooling plays the role
of establishing the level of entropy \citep{voit2001b}, a form of
non--gravitational heating is required to regulate the amount of
lower--entropy gas which is destined to cool and form stars.

As for heating with $K_{\rm fl}=100$ keV cm$^2$ (upper right panel of
Fig. \ref{fig3}), it has a rather small effect on the gas entropy,
while it does have a significant effect on the fraction of stellar
mass, which drops to 10--20 per cent. This result implies that, with
this level of $K_{\rm fl}$, radiative cooling is still the main
responsible for setting the entropy level, while extra heating
regulate the amount of cooled gas. A further increase of $K_{\rm fl}$
is then required to alleviate the tension between the observed and the
measured levels of entropy in galaxy groups. Indeed, using $K_{\rm
  fl}=300$ keV cm$^2$ (bottom panels of Fig. \ref{fig3}) brings the
entropy level in the simulated groups to better match the observed
one. However, as discussed
in the previous section
imposing an entropy floor generates too large voids in the \lya
forest, an effect that can be compensated by increasing the heating
overdensity threshold, $\delta_{\rm h}$. However, increasing the
latter from $\delta_{\rm h}=10$ (bottom left panel) to 30 (bottom
right panel) induces a slight but sizable decrease of $K_\Delta$,
especially for $\Delta=2500$, as a consequence of the smaller number
of gas particles heated at $z=4$. Therefore, while heating at
relatively high overdensity is required by the void statistics of the
\lya forest, it goes in the wrong direction to reproduce the
thermodynamical properties of intra--group medium at small radii. This
demonstrate how effective is combining observations of the high--$z$
IGM and low--$z$ intra--group medium to constrain the thermal history
of the cosmic baryons.

\section{Conclusions}
We presented an analysis of cosmological hydrodynamical simulations,
including radiative cooling and pre--heating, aimed at characterising
both the properties of the high--redshift \lya forest and the
thermodynamical properties of the diffuse gas within nearby galaxy
groups. We use a simple phenomenological recipe for pre--heating, in
which at $z=4$ all the gas particles lying at an overdensity above
$\delta_{\rm h}$ are brought to a minimum entropy level of $K_{\rm
  fl}$. At $z=2.2$ mock QSO spectra were extracted and their
properties compared to the observed distribution of voids in the \lya
flux \citep{viel08voids}, while at $z=0$ the entropy--temperature
relation of galaxy groups is compared with recent results from X--ray
{\small CHANDRA} observations \citep{sun08}. The main results of our
analysis can be summarised as follows: $i)$ pre--heating all the gas,
irrespective of its density, produces voids in the \lya forest, which
are too large if compared to observations; $ii)$ imposing a modest
overdensity threshold for heating, $\delta_{\rm h}=3$, suddenly
reduces the size of the voids to values which are even too small;
$iii)$ further increasing $\delta_{\rm h}$ makes the void sizes
approaching those of the non pre--heated run, owing to the
progressively smaller amount of heated gas which ends up in the \lya
forest; $iv)$ the entropy level within galaxy groups at $r_{500}$ can
be already matched in non pre--heated simulations, which however fail
at accounting for this level at $r_{2500}$; $v)$ a fairly strong
entropy injection, with $K_{\rm fl}>100$ keV cm$^2$, is required to
match the entropy--temperature relation of poor systems at $r_{2500}$.

The need of reproducing the entropy structure of the low--redshift
intra--group medium and the void statistics of the high--redshift \lya
forest leads us to conclude that for a mechanism of non--gravitational
heating to work, it must provide a fairly large amount of extra
entropy at relatively high overdensities, comparable to those
characteristic of virialized halos.  Our conclusion, in agreement with
the argument by \cite{shang}, is that the amount of pre--heating
required by low--redshift galaxy clusters and groups should avoid low
density regions for it not to produce too large voids in the \lya
forest or should act at lower redshift, $z\mincir 1$.  Admittedly, our
scheme of non--gravitational heating is an oversimplified one. For
this reason, our aim is not to look for the best values of $K_{\rm
  fl}$ and $\delta_{\rm h}$, which are able to fit at the same time
the observed intra--group entropy and void statistics of the \lya
forest. Rather, the main goal of our analysis is to provide an
indication of the general properties that a plausible mechanism of
non--gravitational heating should have. We defer to a forthcoming
analysis a detailed study of the combined constraints of the low and
high--$z$ IGM properties using physically motivated models of heating
from astrophysical sources, such as supernovae, active galactic nuclei
and DM annihilation.

\section*{Acknowledgements.}
We acknowledge useful discussions with D. Fabjan, M. Sun and
M. Voit. We thank G. Murante for help with the halo finder. The
simulations were performed with the Darwin Supercomputer at the High
Performance Computing Service of the University of Cambridge
(http://www.hpc.cam.ac.uk/). This work has been partially supported by
the INFN-PD51 grant and by the ASI-AAE Theory Grant.


\bibliographystyle{mn2e} 
\bibliography{master2.bib}

\newcommand{\noopsort}[1]{}
\begin{thebibliography}{}

\bibitem[\protect\citeauthoryear{{Balogh}, {Pearce}, {Bower} \& {Kay}}{{Balogh}
  et~al.}{2001}]{balogh01}
{Balogh} M.~L.,  {Pearce} F.~R.,  {Bower} R.~G.,    {Kay} S.~T.,  2001, \mnras,
  326, 1228

\bibitem[\protect\citeauthoryear{{Bialek}, {Evrard} \& {Mohr}}{{Bialek}
  et~al.}{2001}]{bialek01}
{Bialek} J.~J.,  {Evrard} A.~E.,    {Mohr} J.~J.,  2001, \apj, 555, 597

\bibitem[\protect\citeauthoryear{{Bolton}, {Haehnelt}, {Viel} \&
  {Springel}}{{Bolton} et~al.}{2005}]{bolton05}
{Bolton} J.~S.,  {Haehnelt} M.~G.,  {Viel} M.,    {Springel} V.,  2005, \mnras,
  357, 1178

\bibitem[\protect\citeauthoryear{{Bolton}, {Viel}, {Kim}, {Haehnelt} \&
  {Carswell}}{{Bolton} et~al.}{2008}]{bolton08}
{Bolton} J.~S.,  {Viel} M.,  {Kim} T.-S.,  {Haehnelt} M.~G.,    {Carswell}
  R.~F.,  2008, \mnras, 386, 1131

\bibitem[\protect\citeauthoryear{{Borgani}, {Dolag}, {Murante}, {Cheng},
  {Springel}, {Diaferio}, {Moscardini}, {Tormen}, {Tornatore} \&
  {Tozzi}}{{Borgani} et~al.}{2006}]{borgani06}
{Borgani} S.,  {Dolag} K.,  {Murante} G.,  {Cheng} L.-M.,  {Springel} V.,
  {Diaferio} A.,  {Moscardini} L.,  {Tormen} G.,  {Tornatore} L.,    {Tozzi}
  P.,  2006, \mnras, 367, 1641

\bibitem[\protect\citeauthoryear{{Borgani}, {Fabjan}, {Tornatore}, {Schindler},
  {Dolag} \& {Diaferio}}{{Borgani} et~al.}{2008}]{borgani08}
{Borgani} S.,  {Fabjan} D.,  {Tornatore} L.,  {Schindler} S.,  {Dolag} K.,
  {Diaferio} A.,  2008, Space Science Reviews, 134, 379

\bibitem[\protect\citeauthoryear{{Borgani}, {Finoguenov}, {Kay}, {Ponman},
  {Springel}, {Tozzi} \& {Voit}}{{Borgani} et~al.}{2005}]{borgani05}
{Borgani} S.,  {Finoguenov} A.,  {Kay} S.~T.,  {Ponman} T.~J.,  {Springel} V.,
  {Tozzi} P.,    {Voit} G.~M.,  2005, \mnras, 361, 233

\bibitem[\protect\citeauthoryear{{Gonzalez}, {Zaritsky} \&
  {Zabludoff}}{{Gonzalez} et~al.}{2007}]{gonzalez07}
{Gonzalez} A.~H.,  {Zaritsky} D.,    {Zabludoff} A.~I.,  2007, \apj, 666, 147

\bibitem[\protect\citeauthoryear{{Kay}, {Pearce}, {Jenkins}, {Frenk}, {White},
  {Thomas} \& {Couchman}}{{Kay} et~al.}{2000}]{kay00}
{Kay} S.~T.,  {Pearce} F.~R.,  {Jenkins} A.,  {Frenk} C.~S.,  {White} S.~D.~M.,
   {Thomas} P.~A.,    {Couchman} H.~M.~P.,  2000, \mnras, 316, 374

\bibitem[\protect\citeauthoryear{{Kim}, {Bolton}, {Viel}, {Haehnelt} \&
  {Carswell}}{{Kim} et~al.}{2007}]{tkim}
{Kim} T.~.,  {Bolton} J.~S.,  {Viel} M.,  {Haehnelt} M.~G.,    {Carswell}
  R.~F.,  2007, \mnras, 382, 1657

\bibitem[\protect\citeauthoryear{{Komatsu}}{{Komatsu}}{2008}]{komatsu08}
{Komatsu} E. e.~a.,  2008, ArXiv e-prints, 803

\bibitem[\protect\citeauthoryear{{Lesgourgues}, {Viel}, {Haehnelt} \&
  {Massey}}{{Lesgourgues} et~al.}{2007}]{lesg}
{Lesgourgues} J.,  {Viel} M.,  {Haehnelt} M.~G.,    {Massey} R.,  2007, Journal
  of Cosmology and Astro-Particle Physics, 11, 8

\bibitem[\protect\citeauthoryear{{Mazzotta}, {Rasia}, {Moscardini} \&
  {Tormen}}{{Mazzotta} et~al.}{2004}]{mazzotta2004}
{Mazzotta} P.,  {Rasia} E.,  {Moscardini} L.,    {Tormen} G.,  2004, \mnras,
  354, 10

\bibitem[\protect\citeauthoryear{{Meiksin}}{{Meiksin}}{2007}]{meiksin07}
{Meiksin} A.~A.,  2007, ArXiv e-prints, 711

\bibitem[\protect\citeauthoryear{{Muanwong}, {Thomas}, {Kay} \&
  {Pearce}}{{Muanwong} et~al.}{2002}]{muanwong02}
{Muanwong} O.,  {Thomas} P.~A.,  {Kay} S.~T.,    {Pearce} F.~R.,  2002, \mnras,
  336, 527

\bibitem[\protect\citeauthoryear{{Nagai}, {Kravtsov} \& {Vikhlinin}}{{Nagai}
  et~al.}{2007}]{nagai07}
{Nagai} D.,  {Kravtsov} A.~V.,    {Vikhlinin} A.,  2007, \apj, 668, 1

\bibitem[\protect\citeauthoryear{{Piffaretti}, {Jetzer}, {Kaastra} \&
  {Tamura}}{{Piffaretti} et~al.}{2005}]{piffaretti05}
{Piffaretti} R.,  {Jetzer} P.,  {Kaastra} J.~S.,    {Tamura} T.,  2005, \aap,
  433, 101

\bibitem[\protect\citeauthoryear{{Ponman}, {Sanderson} \&
  {Finoguenov}}{{Ponman} et~al.}{2003}]{ponman03}
{Ponman} T.~J.,  {Sanderson} A.~J.~R.,    {Finoguenov} A.,  2003, \mnras, 343,
  331

\bibitem[\protect\citeauthoryear{{Pratt} \& {Arnaud}}{{Pratt} \&
  {Arnaud}}{2005}]{pratt05}
{Pratt} G.~W.,  {Arnaud} M.,  2005, \aap, 429, 791

\bibitem[\protect\citeauthoryear{{Shang}, {Crotts} \& {Haiman}}{{Shang}
  et~al.}{2007}]{shang}
{Shang} C.,  {Crotts} A.,    {Haiman} Z.,  2007, \apj, 671, 136

\bibitem[\protect\citeauthoryear{{Springel}}{{Springel}}{2005}]{springel}
{Springel} V.,  2005, \mnras, 364, 1105

\bibitem[\protect\citeauthoryear{{Sun}, {Voit}, {Donahue}, {Jones} \&
  {Forman}}{{Sun} et~al.}{2008}]{sun08}
{Sun} M.,  {Voit} G.~M.,  {Donahue} M.,  {Jones} C.,    {Forman} W.,  2008,
  ArXiv e-prints, 805

\bibitem[\protect\citeauthoryear{{Tornatore}, {Borgani}, {Springel},
  {Matteucci}, {Menci} \& {Murante}}{{Tornatore} et~al.}{2003}]{tornatore2003}
{Tornatore} L.,  {Borgani} S.,  {Springel} V.,  {Matteucci} F.,  {Menci} N.,
  {Murante} G.,  2003, \mnras, 342, 1025

\bibitem[\protect\citeauthoryear{{Tozzi} \& {Norman}}{{Tozzi} \&
  {Norman}}{2001}]{TN01}
{Tozzi} P.,  {Norman} C.,  2001, \apj, 546, 63

\bibitem[\protect\citeauthoryear{{Viel}, {Colberg} \& {Kim}}{{Viel}
  et~al.}{2008}]{viel08voids}
{Viel} M.,  {Colberg} J.~M.,    {Kim} T.-S.,  2008, \mnras, 386, 1285

\bibitem[\protect\citeauthoryear{{Viel}, {Haehnelt} \& {Springel}}{{Viel}
  et~al.}{2004}]{viel04}
{Viel} M.,  {Haehnelt} M.~G.,    {Springel} V.,  2004, \mnras, 354, 684

\bibitem[\protect\citeauthoryear{{Vikhlinin}}{{Vikhlinin}}{2006}]{vikhlinin06}
{Vikhlinin} A.,  2006, \apj, 640, 710

\bibitem[\protect\citeauthoryear{{Voit}}{{Voit}}{2005}]{voit2005}
{Voit} G.~M.,  2005, Reviews of Modern Physics, 77, 207

\bibitem[\protect\citeauthoryear{{Voit} \& {Bryan}}{{Voit} \&
  {Bryan}}{2001}]{voit2001b}
{Voit} G.~M.,  {Bryan} G.~L.,  2001, \nat, 414, 425

\bibitem[\protect\citeauthoryear{{Voit}, {Kay} \& {Bryan}}{{Voit}
  et~al.}{2005}]{voit05}
{Voit} G.~M.,  {Kay} S.~T.,    {Bryan} G.~L.,  2005, \mnras, 364, 909

\bibitem[\protect\citeauthoryear{{Younger} \& {Bryan}}{{Younger} \&
  {Bryan}}{2007}]{younger07}
{Younger} J.~D.,  {Bryan} G.~L.,  2007, \apj, 666, 647

\end{thebibliography}

\end{document}